\documentstyle[11pt,newpasp,twoside]{article}
\input{psfig.sty}

\def\edcomment#1{\iffalse\marginpar{\raggedright\sl#1\/}\else\relax\fi}
\reversemarginpar

\begin{document}
\title{Massive  star formation in Luminous Infrared Galaxies}
 \author{Almudena Alonso-Herrero, George H. Rieke, Marcia J. Rieke}
\affil{Steward Observatory, University of Arizona, Tucson, AZ 85721, USA}
\author{Nick Z. Scoville}
\affil{California Institute of Technology, Pasadena, CA 91125, USA}

\begin{abstract}
We present {\it HST}/NICMOS observations of a sample of LIRGs. We show that 
active star formation appears to be occurring
not only in the bright nuclei of these galaxies, 
but also in luminous super-star clusters
and giant H\,{\sc ii} regions 
with ages of up to $20-40\,$Myr. This population of bright
clusters and H\,{\sc ii} regions is unprecedented in 
normal galaxies and emphasizes the effects of the extreme star formation 
in LIRGs.
\end{abstract}

\section{Introduction}

Luminous and ultraluminous infrared galaxies (LIRGs
and ULIRGs, with 
$L_{\rm IR} = 10^{11} - 10^{12}\,$L$_\odot$ and 
$L_{\rm IR} > 10^{12}\,$L$_\odot,$ respectively) have long been
recognized as one of the best laboratories to study the 
process of violent star formation in the Local Universe.
The dust-rich environments of LIRGs and ULIRGs are thought 
to be similar to the conditions in which
star formation 
occurred at high redshift.

The {\it Hubble Space Telescope} (HST) is proving to be an invaluable 
tool for unveiling the  star formation processes in galaxies over spatial 
scales previously unattainable -- scales of a few 
tens of parsecs. Most remarkable is the discovery of the so-called super 
star clusters (SSC) in interacting/merging galaxies (e.g., the Antennae).
Although there is no precise definition, SSCs are  
massive star clusters with luminosities
a few orders of magnitude brighter than globular clusters 
(see a recent review by Whitmore 2000). It is now 
clear that this population of SSCs is not only inherent to interacting 
galaxies, but also to LIRGs (see e.g., Alonso-Herrero et 
al. 2000; 2001; 2002), ULIRGs (Scoville et al. 2000), groups of galaxies 
(Gallagher et al. 2001) and even isolated 
galaxies (e.g., Maoz et al. 2001).

One of the main difficulties in quantifying the age of SSCs
in LIRGs and interacting galaxies is breaking the 
age-extinction degeneracy. This usually translates into only rough 
age estimates for SSCs -- 5 and  900\,Myr, 
from photometric data (Whitmore 2000). 
H\,{\sc ii} regions, on the other hand,  will highlight the youngest 
regions of star formation, with ages of $< 5-10\,$Myr, as 
these are the lifetimes of the O and B
stars required to ionize the gas. In this paper we analyze 
the physical properties of H\,{\sc ii} regions and 
star clusters in a sample of LIRGs,
as well as  their relation and evolution to provide further 
insight into the nature of 
the off-nuclear star formation in LIRGs.

\section{Sample and Observations}
We have selected a sample  of eight LIRGs with both 
{\it HST}/NICMOS narrow-band Pa$\alpha$ ($\lambda_{\rm rest} = 1.87\,\mu$m) 
images and broad-band $H$ 
($1.6\,\mu$m) 
continuum images (Table~1) to identify H\,{\sc ii} regions and 
star clusters, respectively. 

\begin{table}
\caption{Sample of LIRGs.}
\begin{tabular}{cccccc}
\hline
{Galaxy}  &$\log L_{\rm IR}$  &Dist & {FOV of Pa$\alpha$ } &
{$\log L({\rm H}\alpha)_{\rm tot}$} & 
{$L_{\rm nuc}/L_{\rm tot}$} \\ 
 &(L$_\odot$)  &(Mpc) & (kpc$\times$kpc)&{(erg s$^{-1}$)} &  \\
(1)    & (2) & (3) & (4) & (5) & (6)\\
\hline
NGC~6808  & 10.94 & 46 & $11\times 11$   & 41.38 & $\simeq 0$ \\
NGC~5653  & 11.01 & 47 & $11\times 11$   & 41.72 & $\simeq 0$ \\
Zw~049.057& 11.22 & 52 & $4.9\times 4.9$ & 41.21 & -- \\
NGC~3256  & 11.48 & 37 & $3.5\times 3.5$ & 42.11 & 0.41 \\
NGC~1614  & 11.62 & 64 & $6.4\times 6.4$ & 42.60 & 0.65 \\
VV~114    & 11.62 & 80 & $9.3\times 9.3$ & 42.55$^*$ & 0.37 \\
IC~694    & 11.91 & 42 & $3.8\times 3.8$ & 41.95 & 0.58 \\ 
NGC~3690  &       & 42 & $3.8\times 3.8$ & 42.16 & 0.18 \\ 
NGC~6240  & 11.82 & 97 & $9.2\times 9.2$ & 43.19$^*$ & $\simeq 1^*$\\
\hline
\end{tabular}

{\small Notes. --- Column~(1): Galaxy. The two components of Arp~299 are 
usually referred to as IC~694 and NGC~3690.
Column~(2): IR ($8-1000\,\mu$m) luminosity.
Column~(3): Distance. Columns~(4) and (5): Area imaged in Pa$\alpha$   
and H$\alpha$ luminosity over that area. 
Column~(6): Ratio of
the nuclear to total H$\alpha$ luminosity. 
$^*$ Uncertain because of the large 
correction needed to account for the total  
Pa$\alpha$ flux (see AAH02 for details).}
\end{table}

\section{Super Star Clusters}

Much of the recent star formation in our sample of LIRGs 
appears to be occurring
not only in the bright nuclei, but also in luminous clusters 
and H\,{\sc ii} regions (e.g., Table~1, last column), similar to those 
found in other
interacting and highly luminous IR galaxies (e.g., 
Scoville et al. 2000; AAH02). The absolute $H$-band 
magnitudes for clusters in LIRGs (not corrected for extinction) range up 
to approximately $M_H =-17\,$mag to $M_H =-18\,$mag (see histogram for 
the clusters of NGC~3256 in Fig.~1, left panel). The lower detection limit 
depends on the emission from the underlying galaxy and the degree 
of crowding. For 
instance the distribution of $H$-band luminosities of clusters
detected in NGC~3256 appears 
to be complete down to $M_H\simeq -14\,$mag (Fig.~1).

The luminosities of 
the brightest IR clusters 
in LIRGs may exceed the limits found in more 
normal conditions. For example, the intermediate-age clusters in
M100 (Ryder \& Knapen 1999) have $M_H = -12\,$mag to $M_H = -15\,$mag 
assuming ($H-K \simeq 0.2$). Since the IR luminosities change only slowly with 
time after approximately 20 million years (see Fig.~2),
we can compare directly to see that the luminosities of clusters in normal 
galaxies may be about $1.5-2\,$mag lower than 
for LIRGs. Even when compared to starburst galaxies, LIRGs appear to 
have an excess of luminous 
IR clusters, as illustrated in Fig.~1 (left panels). This
figure compares the distribution of luminosities of clusters 
in the central region of 
the starburst galaxy NGC~1530 (at a distance similar to  NGC~3256 and thus 
same spatial resolution) with those detected in NGC~3256.

\begin{figure}[h]
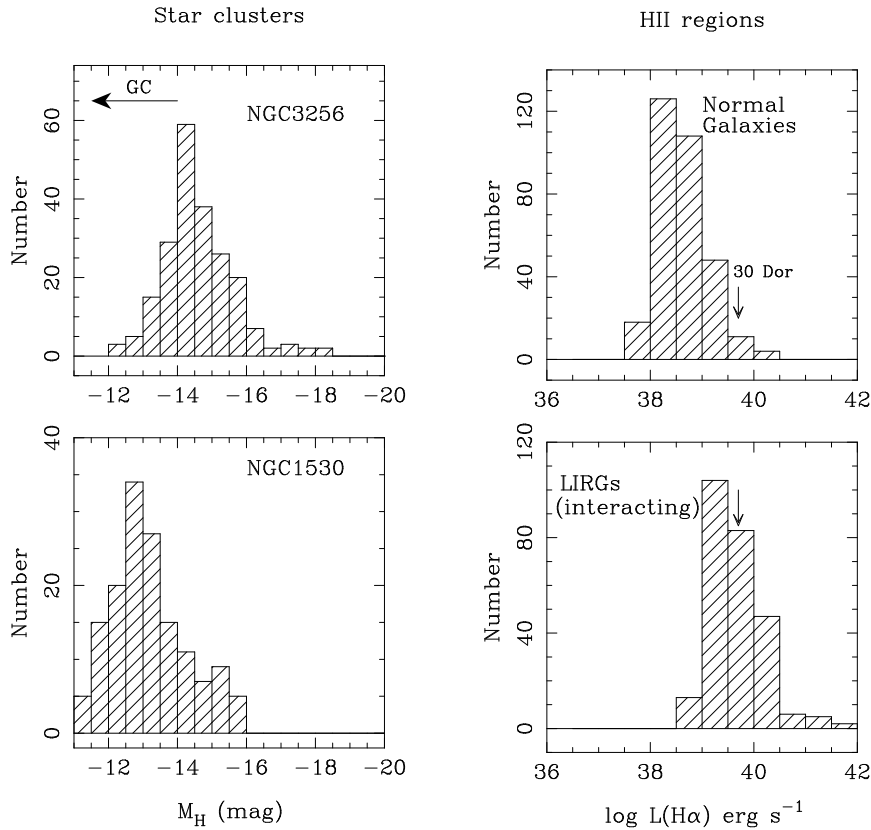

\plotfiddle{aalonso_fig1a.ps}{425pt}{-90}{65}{65}{-195}{490}
\plotfiddle{aalonso_fig1b.ps}{425pt}{-90}{65}{65}{-15}{925}
\vspace{-20cm}
\caption{{\it Left panels:} comparison of the absolute
$H$-band magnitudes (from {\it HST}/NICMOS observations) 
of star clusters detected in 
the LIRG NGC~3256 and the starburst galaxy NGC~1530. We also 
show the typical magnitudes of globular clusters (see e.g., 
Kissler-Patig et al. 2002). {\it Right panels:} 
H$\alpha$ luminosities of 
H\,{\sc ii} regions detected in LIRGs (NGC~3256 and Arp~299)
and in normal galaxies. The arrow indicates the luminosity of 30
Dor, the prototypical giant H\,{\sc ii} region.}
\end{figure}

\section{Giant H\,{\sc ii} Regions}

In two previous studies we showed the presence of 
a population of bright H\,{\sc ii} regions in two LIRGs, 
Arp~299 (AAH00) and NGC~1614 (AAH01). A significant fraction of
these H\,{\sc ii} regions displays H$\alpha$ 
luminosities in excess of that of 
30 Doradus, the prototypical giant H\,{\sc ii} region. The analysis
of the sample of LIRGs in Table~1 has revealed 
that giant H\,{\sc ii} regions are 
ubiquitous in LIRGs and are located not only in and near the nuclei of 
interacting galaxies, but also at the
interface of interacting galaxies and along the spiral arms of
isolated systems.

In Fig.~1 (right panels) we compare the 
H$\alpha$ luminosities (not corrected for
extinction)  
of H\,{\sc ii} regions in LIRGs with those in a small sample of normal 
galaxies observed with the same spatial resolution 
from Alonso-Herrero \& Knapen (2001). Giant H\,{\sc ii} regions 
(the luminosity of 30 Doradus is indicated with an arrow in Fig.~1) 
are more common in LIRGs 
than in normal galaxies. The measured sizes of giant H\,{\sc ii} 
regions in LIRGs when compared to those of normal galaxies rule out the 
possibility that these giant H\,{\sc ii} regions are just aggregations of 
"normal" H\,{\sc ii} regions. A more plausible explanation for this population 
of luminous H\,{\sc ii} regions in LIRGs is that regions of high gas 
pressure and density in LIRGs, ULIRGs, and interacting 
galaxies provide the
necessary conditions for the formation of a large number of 
massive star (ionizing) clusters. Such extreme conditions 
are not likely to occur in normal galaxies. 

\begin{figure}
\plotfiddle{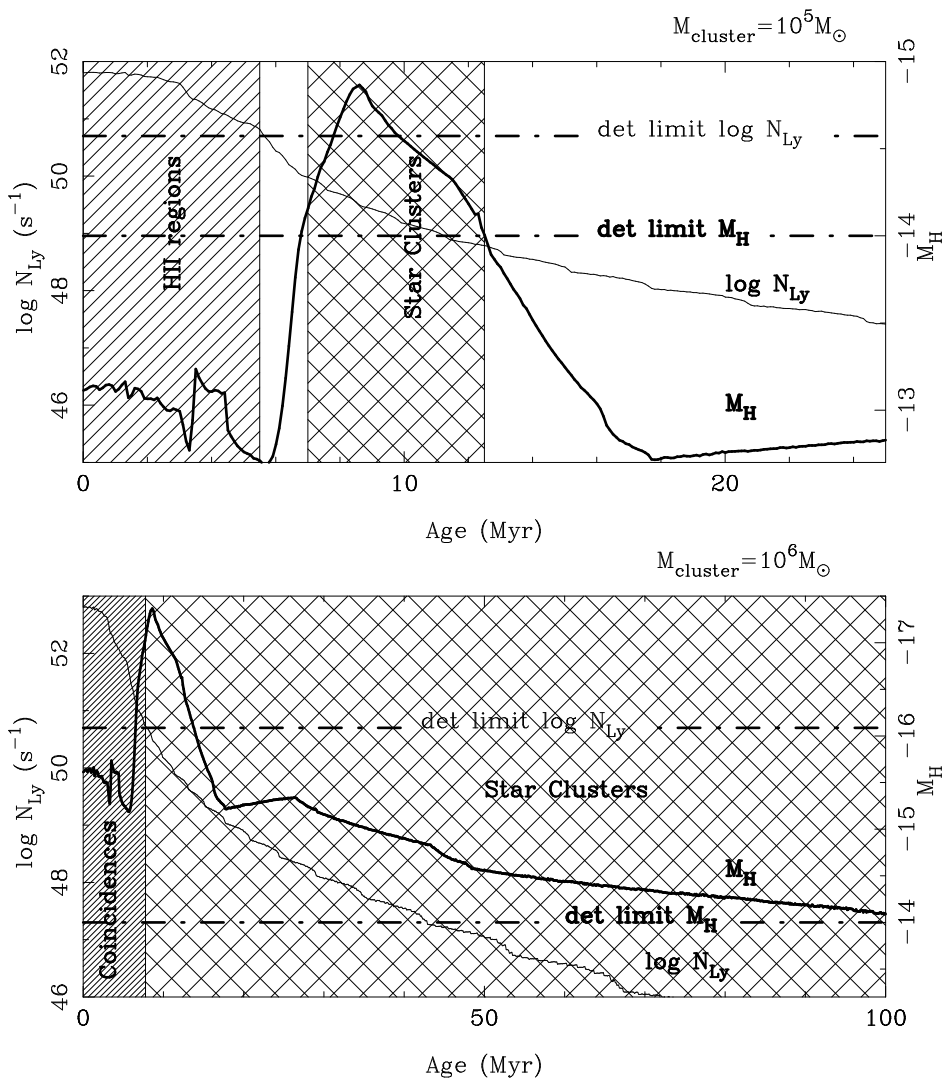}{425pt}{0}{70}{70}{-220}{0}
\vspace{0.cm}
\caption{The solid lines are outputs of Starburst99 (Leitherer et al. 
1999) showing the time evolution of the number of ionizing photons 
($\log N_{\rm Ly}$, thin solid line, scale on the left hand side) 
and absolute $H$-band magnitude ($M_H$,
thick solid line, scale on the right hand side) 
for a $10^6\,{\rm M}_\odot$ cluster (bottom panel) 
and a $10^5\,{\rm M}_\odot$ cluster (top panel), for 
an instantaneous burst, a Salpeter IMF 
(between 1 and $100\,{\rm M}_\odot$) and solar metallicity.
The dashed-dotted lines represent the approximate 
detection thresholds for NGC~3256: $M_H=-14\,$mag and $\log N_{\rm Ly}
=50.7\,{\rm s}^{-1}$. An H\,{\sc ii} region with no cluster counterpart will 
be the case when $\log N_{\rm Ly}$ is above the detection threshold, but 
$M_H$ is not detectable yet (lightly hatched 
area). There will be a coincidence between 
an H\,{\sc ii} region and a near-IR star cluster when 
both $N_{\rm Ly}$ and $M_H$ are above their 
respectively detection thresholds (closely hatched 
area). We will only observe a star cluster when $\log N_{\rm Ly} < 
50.7\,{\rm s}^{-1}$, but $M_H$ is still observable (cross-hatched area).}
\end{figure}

\section{Giant H\,{\sc ii} Regions and their relation to SSC: the 
age sequence}

Despite the large numbers of near-IR SSCs and H\,{\sc ii} 
regions identified 
in LIRGs, there is only a small fraction of coincidences
($4-30\,$\%)  between H\,{\sc ii} regions and star 
clusters. We can use evolutionary synthesis models to reproduce the 
observed relative fractions of young and intermediate H\,{\sc ii} regions 
or clusters and old clusters in Arp~299 and NGC~3256. 
In Fig.~2 we show outputs of Starburst99 (Leitherer et al. 
1999) for the time evolution of the absolute $H$-band magnitude 
and number of ionizing photons. We show two cluster masses and 
instantaneous star formation with a Salpeter IMF. 
For these assumptions and taking into account 
the detection limits for the complete distributions of
$H$-band luminosities of clusters 
in NGC~3256 and Arp~299 we infer 
photometric masses for the detected clusters of between $\simeq 5 \times 10^4$ 
and $10^6\,{\rm M}_\odot$. 

The fact that the peak of the $H$-band 
luminosity occurs after 
approximately 9\,Myr, whereas at the same time the 
number of ionizing photons has
dropped by about 2 orders of magnitude from the maximum, provides an 
explanation for
the limited number of coincidences. Within the present detection 
limits in Arp~299
and NGC~3256, we can detect both H\,{\sc ii} region emission and a star 
cluster for the most massive clusters ($\simeq 10^6\,{\rm M}_\odot$) 
only during the first 7\,Myr (Fig.~2). The
near-IR clusters with no detected H\,{\sc ii} region emission will 
be older than approximately 7\,Myr. The H\,{\sc ii} regions with no 
detected cluster
counterpart are most likely younger than 5\,Myr, and have 
intermediate-mass ($5 \times 10^4-10^5\,{\rm M}_\odot$) 
ionizing clusters. If, as observed in obscured
Galactic H\,{\sc ii} regions, there are significant amounts of 
extinction during the first million years of the evolution 
of clusters and associated H\,{\sc ii} regions, then the 
observed fractions of H\,{\sc ii}
 regions and coincidences will be lower limits.

     An estimate of the age distribution of the observed 
clusters can be inferred from the relative numbers of H\,{\sc ii} 
regions and near-IR star
clusters and the model predictions: 
The higher the fraction of near-IR clusters compared
to that of H\,{\sc ii} regions, 
the older the ages of the detected star clusters will be. 
The ages of the detected
star clusters in Arp~299 and NGC~3256 range up to $20-40\,$Myr. 
Older clusters possibly created in this or previous episodes 
of star formation are likely
to exist in these systems but cannot be identified with the 
present detection threshold. Another possibility to explain 
the apparent youth of the clusters
in Arp~299 and NGC~3256 would be destruction of clusters. 
In that case, if the clusters have been created at a constant 
rate for the last 100\,Myr,
then roughly 50\% of the clusters are destroyed during that 
time to account for the observed fraction of clusters in 
these two systems. The data
presented in this paper does not allow us to distinguish 
between these two possibilities. 

     From the present observations and modeling we find that a large 
fraction of the youngest clusters (that is, the ionizing clusters of the 
H\,{\sc ii}
regions with ages less than $5-6\,$Myr) will not be detected 
from near-IR continuum imaging alone, as only some 
8\%--16\% of these H\,{\sc ii} regions in
our sample of LIRGs  
appear to have near-IR cluster counterparts. This suggests that studies
of the young star clusters in galaxies performed using only 
near-IR continuum imaging may be missing a significant fraction of the 
youngest star-forming
regions.

\section*{Acknowledgments}
AAH participation in this conference was made possible by a
travel grant from the AAS.  The AAS travel
grant program is supported by the National Science Foundation.

\section*{Discussion}

\noindent
{\it J. M. Mas-Hesse:} The peak in $H$-band emission predicted 
by evolutionary synthesis models at around 10\,Myr is due to the 
formation of Red Supergiants. Predictions for RSGs are very much 
model-dependent, since they are strongly affected by rotation, 
so that they have to be taken with care.\\
\noindent
{\it A. Alonso-Herrero:} Yes, I'm aware of this problem 
and obviously some of the results I've presented are model-dependent. \\
\noindent
{\it G. Tenorio-Tagle:} Can you please indicate the physical
size of the SSCs.\\
\noindent
{\it A. Alonso-Herrero:} The sizes of the super star clusters
are somewhat dependent on the spatial resolution (i.e., the 
distance of the galaxy). For the closest galaxies in our 
sample the typical diameters are of the order of $20-30\,$pc. \\

\end{document}